\documentclass[12pt]{article}
\usepackage{xspace}
\usepackage{lutypaper}
\usepackage{mydefs}
\usepackage{equations}

\newcommand{\fund}{\drawsquare{6.5}{0.4}}
\newcommand{\afund}{\overline{\fund}}

\newcommand{\asymm}{\raisebox{-3pt}{\drawsquare{6.5}{0.4}\hskip-6.9pt%
        \raisebox{6.5pt}{\drawsquare{6.5}{0.4}}}}

\newcommand{\asymmthree}{\raisebox{-7pt}{\drawsquare{6.5}{0.4}}\hskip-6.9pt%
\raisebox{-0.5pt}{\drawsquare{6.5}{0.4}}\hskip-6.9pt%
\raisebox{6pt}{\drawsquare{6.5}{0.4}}}

\renewcommand{\d}{\partial}

\newcommand{\bb}[1]{\hbox{\bf #1}}

\begin{document}

\begin{titlepage}

\vskip -.4in
\preprint{UMDHEP 99-68\\
UCB-PTH-98/60,
LBNL-42601}

\title{Improved Single Sector Supersymmetry Breaking}

\author{Markus A. Luty\footnote{Sloan Fellow}}

\address{Department of Physics,
University of Maryland\\
College Park, Maryland 20742, USA\\
{\tt mluty@physics.umd.edu}}

\author{John Terning}

\address{Department of Physics,
University of California\\
and\\
Theory Group, Lawrence Berkeley Laboratory\\
Berkeley, California 94720, USA\\
{\tt terning@alvin.lbl.gov}}

\begin{abstract}
Building on recent work by N. Arkani-Hamed and the present authors, we
construct realistic models that break supersymmetry dynamically and give
rise to composite quarks and leptons, all in a single strongly-coupled
sector.  The most important improvement compared to earlier models is that
the second-generation composite states correspond to dimension-2 "meson"
operators in the ultraviolet.  This leads to a higher scale for flavor
physics, and gives a completely natural suppression of flavor-changing
neutral currents.  We also construct models in which the hierarchy of
Yukawa couplings is explained by the dimensionality of composite states.
These models provide an interesting and viable alternative to gravity- and
gauge-mediated models.  The generic signatures are unification of scalar
masses with different quantum numbers at the compositeness scale, and
lighter gaugino, Higgsino, and third-generation squark and slepton masses.
We also analyze large classes of models that give rise to both
compositeness and supersymmetry breaking, based on gauge theories with
confining, fixed-point, or free-magnetic dynamics.
\end{abstract}


\end{titlepage}

\section{Introduction}
One of the most exciting results of the recent progress in
understanding non-pertur\-bative effects in
supersymmetric gauge theories \cite{Seiberg} is that it
allows the exploration of new possibilities for the realization of
supersymmetry (SUSY) in nature.
The most important example is the use of dynamical SUSY
breaking to explain the origin of the SUSY breaking scale.
In recent years, large classes of SUSY
gauge theories have been discovered
that exhibit dynamical SUSY breaking through a
variety of different mechanisms \cite{break},
and realistic models have been built using these theories
as building blocks in both (super)gravity-mediated \cite{gravmed}
and gauge-mediated frameworks \cite{GaMed}.

In these conventional approaches to SUSY model building, SUSY breaking
arises in a separate sector consisting of fields that are neutral under the
standard model gauge group, and SUSY breaking is communicated to the
observable fields by messenger (gauge or gravitational) interactions.
It is clearly important to know whether such a `modular' structure is required
in order for SUSY to be the solution of the hierarchy problem,
or if simpler models without sectors are possible.
In \Ref{ALT}, realistic models were constructed that do not
require a separate SUSY breaking sector.
In these models, SUSY is broken dynamically by fields that are charged
under the standard model gauge group, giving rise to composite
fields with the quantum numbers of quarks and leptons.
(Compositeness avoids the `no-go' theorem of Dimopoulos
and Georgi on SUSY breaking by charged fields \cite{DG}.)
The scalar components of the composite quarks and leptons have
SUSY breaking masses induced directly by the strong dynamics, while
the masses of the fermions are protected by unbroken chiral symmetries.
The masses of the gauginos (which are elementary in this class of models)
arise at 1-loop order, are therefore smaller than the
composite scalar masses, which must therefore be in the range of
1--10 TeV.
Any elementary sfermions in the model also obtain their mass from gauge
mediation from the composite scalars.
There are non-trivial constraints on this scenario arising from the
requirement that the spectrum is phenomenologically acceptable, and
these will be discussed below.

If we make the simplest assumption that the first two generations are
composite while the third is elementary, we automatically gain a partial
understanding of the observed hierarchy of fermion masses.
The reason is that all Yukawa couplings involving composite states
must arise from higher-dimension operators in the fundamental theory,
and are necessarily suppressed, while the top Yukawa coupling can
be order one.%
\footnote{This is true as long as there are no trilinear Yukawa couplings
generated by the strong dynamics, for a model with such dynamical
couplings, but no SUSY breaking see \cite{NelsonStrass}.
For other composite SUSY models, see \Refs{lm,ModComp}.}
A highly non-trivial feature of this scenario is that it does not lead
to excessive flavor-changing neutral currents (FCNC's) from squark
non-degeneracy even if the flavor sector has no flavor symmetry.
This is because the strong composite dynamics is flavor-blind,
and so the composite scalar masses are degenerate to high accuracy,
with small corrections due to perturbative flavor-breaking couplings.%
\footnote{This mechanism is similar to the one that operates in QCD to
give rise to an approximate flavor symmetry for hadrons made of light
quarks.}

We can also consider `dimensional hierarchy'
models in which the first and second generations are composites with
different dimensionality.%
\footnote{Interesting models with this feature were constructed in
\Ref{KLS}.
However, these models do not incorporate SUSY breaking in the manner
envisioned here.}
In these models, there is no symmetry enforcing degeneracy of the
squark masses, and we must assume that squark masses are in the $10\TeV$
range to suppress FCNC's.

These scenarios for single sector SUSY breaking have several
interesting generic phenomenological implications. First, as mentioned
above, gaugino and stop masses will be much smaller
than the masses of composite squarks and sleptons.
Second, the composite scalar masses unify at the compositeness
scale.
(In models where the first two generations are composite, \emph{all} the
scalars of the first generations unify;
in the dimensional hierarchy models the scalars in the first and second
generation unify separately.)
Third, if we assume that the Yukawa interactions are generated by new
physics at a flavor scale above the compositeness scale without special
flavor symmetries, predictions for flavor-changing processes such as
$\mu \to e\ga$ are plausibly within experimental reach.

In this paper, we extend the work of \Ref{ALT} in two important ways. First,
we construct models in which the composite quarks and leptons
correspond to dimension 2 `meson' operators in the fundamental theory
(rather than dimension 3 as in \Ref{ALT}).
This means that the scale of flavor physics is higher in the
new models, leading to a completely natural suppression of FCNC effects,
including $\ep_K$.
(In fact, the desirability of models with dimension-2 composites was
emphasized in \Ref{ALT}.)
Second, we construct explicit dimensional hierarchy models that give a
framework for understanding the fermion mass hierarchy that is closely
linked to the mechanism of SUSY breaking.
Finally, we construct and analyze large classes of supersymmetric gauge
theories with non-perturbative dynamics of the type required for this
kind of model-building.
At low energies the models either
confine (like the models of \Ref{ALT}),
have conformal fixed points,
or are magnetically free.
This shows that the combination of compositeness and SUSY breaking is
not uncommon,
and suggests that further exploration of the connection
between these phenomena is worthwhile.

This paper is organized as follows.
In Section 2, we describe the origin of mass scales and the
generic phenomenology of the models that
we are describing.
In Section 3, we analyze some specific models, and Section 4 contains
our conclusions.
The detailed analysis of specific gauge theories is relegated
to an appendix.

\section{Mass Scales and Phenomenology}
In this Section, we describe the most important qualitative features
of the models constructed in this paper.
Much of this material appears already in \Ref{ALT}, but we present it here
specialized to the two new types of models we will construct:
`meson' models where the first two generations correspond to dimension 2
operators, and `dimensional hierarchy' models in which the first generation
corresponds to dimension 3, the second to dimension 2, and the third
generation is elementary (dimension 1).
We want to emphasize the fact that the phenomenology is very rich, and is
largely independent of the details of specific models.
More detail can be found in \Ref{ALT} and in the appendix to this paper.

\subsection{SUSY Breaking and Compositeness}
\label{SBC}
We first explain the mechanism that gives rise to SUSY breaking and
compositeness.
The models we describe have a strong gauge group of the form
$G_{\rm comp} \times G_{\rm lift}$, where both groups are asymptotically
free%
\footnote{Actually $G_{\rm lift}$ need not be asymptotically free
\cite{ALT}, but we will ignore that possibility here for simplicity.}
and $\La_{\rm comp} \gg \La_{\rm lift}$.
The scale $\La_{\rm comp}$ is the compositeness scale, in the sense that the
degrees of freedom that correspond to the quarks and leptons at low energies
are strongly interacting at the scale $\La_{\rm comp}$.
Direct bounds on the compositeness scale imply that
$\La_{\rm comp} \gsim 2\TeV$.
The role of the gauge group $G_{\rm lift}$ is to generate a dynamical
superpotential that lifts the vacuum degeneracy and gives rise to a local
SUSY breaking minimum.

The models contain the following fields%
\footnote{The names of the fields originate from the fact that these
models are all distant cousins of the venerable `3--2' model of dynamical
SUSY breaking \cite{ADStoo,LutyTerning}.}
\beq\nonumber
\begin{tabular}{c|cc|cc}
& $G_{\rm lift}$ & $G_{\rm comp}$  & $G_{\rm global}$ \\
\hline
$Q$ & \fund & \fund & {\bf 1} \\
$L$ &  $\afund$  & {\bf 1}  & $\overline{\fund}$ \\
$\bar{U}$  & {\bf 1} & $\overline{\fund}$ & \fund \\
$P$  & {\bf 1} & $R$ & {\bf 1} \\
\end{tabular}
\eeq
where the representation $R$ may be highly reducible (implying additional
global symmetries).
In addition, the model has a tree-level superpotential
\beq\eql{lambdadef}
W = \la Q L \bar{U}.
\eeq
There are additional requirements on the model in order for this model
to have a local SUSY breaking minimum.
We choose $G_{\rm global}$ such that classically there is a flat
direction with $\bar{U} \ne 0$ where $Q$ and
$L$ are massive and $G_{\rm comp}$ is completely broken.
Nonperturbative $G_{\rm lift}$ dynamics lift this flat direction via a
dynamical superpotential of the form%
%
%
\beq\eql{wdyn}
W_{\rm dyn} \sim \La_{\rm lift}^{3 - r} \bar{U}^r.
\eeq
Whether this superpotential forces $\bar{U}$ to large or small values depends
on the value of $r$, but it also depends on the effective \Kahler potential
for $\bar{U}$. For large $\bar{U}$,  $G_{\rm comp}$ is completely perturbative
and the \Kahler potential is smooth in
$\bar{U}$, so the potential slopes toward the origin for $r > 1$.%
\footnote{The case $r = 1$ is marginal;
perturbative interactions determine whether the potential slopes toward
or away from the origin \cite{ALT,inverted}.}
For $\bar{U} \ll \La_{\rm comp}$ the $G_{\rm comp}$ dynamics
changes the \Kahler potential for $\bar{U}$.
For example, if the $G_{\rm comp}$ dynamics is confining, the \Kahler
potential will be smooth in terms of a `composite' field $B=(\bar{U}^n)$.
The superpotential can then be written
\beq
W_{\rm dyn} \sim  B^{r/n},
\eeq
which corresponds to a potential that
slopes \emph{away} from the origin if $r/n < 1$.
Therefore, for $1 < r < n$ there is no SUSY minimum for any value of
$\bar{U}$, and there is a SUSY breaking minimum near the border between
the region of validity of the confined and Higgs descriptions.
This occurs for
\beq\eql{uvev}
\avg{\bar{U}} \sim \frac{\sqrt{N} \La_{\rm comp}}{4\pi},
\eeq
where $N$ is the number of `colors' of $G_{\rm comp}$.  For an explanation
of the factors of $4\pi$ and $N$ see \Refs{ALT,fourpiSUSY}.
We keep track of powers of $N$ in our estimates because we will see that
$N \sim 10$ for realistic models.

This mechanism also occurs in the case where the $G_{\rm comp}$ dynamics
gives rise to a conformal fixed point (in the limit where we turn off
$G_{\rm lift}$), provided that the $\bar{U}$ anomalous dimensions are
sufficiently large.
As long as $\bar{U} \ll \La_{\rm comp}$ the $G_{\rm comp}$ dynamics is
controlled by the infrared fixed point.
Recall that we are assuming that $G_{\rm lift}$ is weak at the scale
$\La_{\rm comp}$, so the non-perturbative superpotential can be viewed
as a perturbation.
The 1PI potential for $\bar{U}$ is therefore
\beq\eql{conformV}
V_{\rm 1PI} \simeq (K_{\rm 1PI}^{-1})_{\bar{U}^\dagger\bar{U}} \left|
\frac{\d W_{\rm dyn}}{\d\bar{U}} \right|^2,
\eeq
where $K_{\rm 1PI}$ is the 1PI \Kahler metric evaluated at the conformal
fixed point.
The scaling dimension of the \Kahler metric
$(K_{\rm 1PI})_{\bar{U}^\dagger\bar{U}}$
is $2 - 2 d_{\bar{U}}$, where
$d_{\bar{U}}$ is the scaling dimension of $\bar{U}$.
Therefore,
\beq
(K_{\rm 1PI}^{-1})_{\bar{U}^\dagger\bar{U}}
\sim \bar{U}^{2 (d_{\bar{U}} - 1) / d_{\bar{U}}}.
\eeq
This forces the potential to slope away from the origin for
\beq
\frac{1 - d_{\bar{U}}}{d_{\bar{U}}} > r - 1.
\eeq

One might worry that this argument relies on a `Higgs'
description in terms of the elementary field $\bar{U}$ in a regime where
the theory is strongly coupled.
In many cases there is an alternate description in terms of a weakly-coupled
dual theory.
For example, if $G_{\rm comp} = SU(N)$ with $F$ `flavors' $U, \bar{U}$,
the theory has an infrared fixed point for $\sfrac{3}{2} N < F < 3N$
\cite{SeibergDual}.
There is a dual description in terms of a
theory with gauge group $SU(F - N)$ in which the `baryon'
operator $\bar{U}^N$ in the original
theory is mapped to an operator $\bar{u}^{F - N}$ in the dual.
For $F$ near $\sfrac{3}{2} N$ the dual description is weakly coupled,
and the considerations of the previous paragraph can be made
rigorous.%
%
%
One finds that the behavior of the \Kahler potential agrees precisely
with \Eq{conformV}.
This equivalence between the `Higgs' and `dual' descriptions can be
viewed as a generalization of the usual `complementarity'   \cite{complement}
for theories with scalars in the fundamental representation, and gives
us additional confidence in the considerations above.


We see that there is a general mechanism that can stabilize the field
$\bar{U}$ and give rise to a SUSY-breaking vacuum.
In the models we construct, the above discussion holds only on one branch
of the moduli space, and there are other branches with SUSY minima.
However, the mechanism still gives rise to a metastable \emph{local} SUSY
breaking minimum, which is sufficient.

In these models, SUSY is broken by
\beq
\avg{F_{\bar{U}}} \sim \Avg{\frac{\d W_{\rm dyn}}{\d\bar{U}}}
\sim \frac{\La_{\rm comp}^2}{4\pi} ( \la \sqrt{N} )^{r - 1}
\left( \frac{\La_{\rm lift}}{\La_{\rm comp}} \right)^{3 - r},
\eeq
where we used \Eq{uvev}.
Since $r < 3$ (otherwise the dynamical superpotential \Eq{wdyn}
does not have a good limit
$\La_{\rm lift} \to 0$ when $G_{\rm lift}$ is asymptotically free),
we have $\avg{F_{\bar{U}}} \ll \La_{\rm comp}^2$.%
%
%
The scalar components of $\bar{U}$ get a SUSY-breaking mass of order
\beq
m^2_{\bar{U}} \sim \Avg{\frac{\d^2}{\d\bar{U}^2} \left|
\frac{\d W_{\rm dyn}}{\d\bar{U}} \right|^2}
\sim \frac{F_{\bar{U}}^2}{\avg{\bar{U}}^2} \equiv m^2_{\rm comp}.
\eeq

The `preon' fields $P$ charged under
$G_{\rm comp}$ get SUSY-breaking masses from effects such as
\beq
\Ga_{\rm 1PI} \sim \myint d^2\th d^2\bar{\th}\,
\frac{16\pi^2}{\La_{\rm comp}^2} \bar{U}^\dagger \bar{U} P^\dagger P
\sim m_{\rm comp}^2 P^\dagger P.
\eeq
The fermion components of $P$ can remain massless
because of unbroken chiral
symmetries \cite{tHooft}, and these can be identified with quarks and leptons.

With this discussion of the mass scales, we have
enough information to analyze the main features of the phenomenology of these
models.
Masses for standard-model gauginos and elementary charged scalars are
generated by gauge mediation from the composite scalars, so that
\beq
m_{\la,{\rm SM}} \sim N \frac{g_{\rm SM}^2}{16\pi^2} m_{\rm comp},
\qquad
m^2_{\phi,{\rm elem}} \sim N \left(
\frac{g_{\rm SM}^2}{16\pi^2} m_{\rm comp} \right)^2.
\eeq
Note that the multiplicity factor $N$ enhances gaugino masses compared to
elementary scalar masses.

In the models we construct, some or all of the quarks and leptons from
the first two generations are composite, while the third generation is
elementary.
The reason for this is that in our models
the Yukawa couplings for
composite quarks and leptons arise from higher-dimension operators in the
fundamental theory, and are naturally small compared to one.
It is difficult to accommodate the order-one top Yukawa coupling in this
framework unless the top quark is elementary.
Another reason for the third generation to be elementary is that stop
masses of order $m_{\rm comp} \sim 1$--$10\TeV$ (needed to get sufficiently
large gaugino masses) necessitate a large amount
of fine-tuning in electroweak symmetry breaking.
%
%
These arguments do not forbid the possibility that the right-handed bottom
quark or third-generation leptons are
composite, but we will not take advantage of these loopholes.
In order to obtain a third generation scalar mass
$m_{3} \gsim 100\GeV$ we therefore require
\beq
m_{\rm comp} \gsim \frac{10\TeV}{\sqrt{N}}.
\eeq

We see that
this class of models naturally has a superpartner spectrum similar to the
`more minimal' framework \cite{CoKaNe}.
In models of this kind, there is a dangerous negative contribution to the
third-generation squark masses from the heavy scalars \cite{NimaHitoshi}, given
by
\beq\eql{mrg}
\mu \frac{d m_3^2}{d\mu} = \frac{8 g^2}{16\pi^2} C_2
\left[ \frac{3 g^2}{16\pi^2} m_{\rm comp}^2 - m_\la^2 \right],
\eeq
where we have assumed that a single gauge group dominates and specialized
to the case of two full composite generations.
One way to avoid this problem is to have the compositeness scale close
to $10\TeV$, so that the negative contribution above does not dominate.
If the compositeness scale is high, one can avoid problems if the gaugino
contribution is important. From \Eq{mrg}, we see that
$m_{\la} \gsim m_{\rm comp} / 10$ is sufficient.
(This estimate is confirmed by the detailed analysis of \Ref{NimaHitoshi}.)
This condition is plausibly satisfied if $N \gsim 10$.
In addition, we will see below that the sector that breaks flavor symmetries
and generates Yukawa couplings can plausibly give large positive contributions
to the third-generation scalar masses large enough to eliminate this problem.

Most of the models we construct have of order $N$ `preonic' generations above
the
compositeness scale, and for $N \gsim 10$ the standard-model gauge groups
are far from asymptotically free.
This is compatible with perturbative unification if the compositeness scale
is above (or near) the GUT scale $10^{16}\GeV$.
(Note that for such large compositeness scales, the composite dynamics
need not conserve baryon number.)

In this framework, we automatically gain a partial understanding of
the fermion mass hierarchy:
the Yukawa couplings of the first two generations are naturally suppressed
because they arise from higher-dimension operators in the fundamental theory.
(Since the compositeness scale can be above the GUT scale, an intriguing
possibility is that the flavor scale is the Planck scale.)
If all composite states correspond to operators of the same dimension, there
are further hierarchies in the fermion masses that must be explained;
on the other hand, we will see that a high degree of squark degeneracy can be
guaranteed by the strong dynamics in this case.
We can also consider models in which the composite
states of different generations correspond to operators of different dimension;
in this case, there is no squark degeneracy, and FCNC's must be suppressed
by large squark masses.
We will consider both types of models in what follows.

\subsection{`Meson' Models}
We first discuss `meson' models where all quarks and leptons of the first two
generations correspond to dimension-2 operators $ P \bar{U}$
in the fundamental theory.
This means that Yukawa couplings involving the first two generations can be
generated by adding the following terms to the tree-level superpotential:
\beq\eql{massop}
\De W =
\frac{1}{M^2} H (P \bar{U}) (P \bar{U})
+ \frac{1}{M} H \Phi_3 (P \bar{U}).
\eeq
Here $H$ is a Higgs field and $\Phi_3$ is an elementary third-generation
quark or lepton field.
which gives a Yukawa matrix of the form
\beq
y \sim \pmatrix{
\ep^2 & \ep^2 & \ep \cr
\ep^2 & \ep^2 & \ep \cr
\ep   & \ep   & 1   \cr},
\qquad
\ep \sim \frac{\avg{\bar{U}}}{M},
\eeq
where $M$ is the scale of new physics where flavor symmetries are broken.
Additional structure is clearly needed to construct fully realistic
Yukawa matrices, but for $\ep$ in the range $10^{-1}$--$10^{-2}$
this is a good starting point.

In the above we have assumed that the $G_{\rm comp}$
gauge coupling is perturbative just above the scale $\La_{\rm comp}$
(as in QCD).
This may not be true in models where
the $G_{\rm comp}$ interactions by themselves have an strongly-coupled
infrared fixed point.
In such theories the large anomalous dimension
of $\bar{U}$ can persist up to momentum scales significantly
above the strong dynamics scale.
If this is the case then the operators in \Eq{massop} are enhanced
(just as in `walking technicolor' theories \cite{walking}),
and the flavor scale can be put even higher.

We will make the conservative assumption that the new physics at the scale
$M$ does not have any approximate flavor symmetries that can suppress FCNC's.
In particular, this means that the Yukawa couplings $\la$ in \Eq{lambdadef}
do not conserve flavor.
It is highly non-trivial that the strong dynamics in this theory nevertheless
gives rise to an approximate flavor symmetry at low energies that enforces
the near degeneracy of the composite scalars.
The underlying reason for this is the fact that all of the composites
$(P \bar{U})$ are part of a single multiplet from the point of view of the
strong interactions.

Let us first consider the $\la$-dependent effects.
The superpotential \Eq{wdyn} depends on $\la$ only through
$\det(\la)$, which is flavor independent.
There is nontrivial $\la$ dependence in the effective \Kahler potential,
but it is proportional to $\la^2 / (16\pi^2) \lsim 10^{-2}$.
We now consider the effects of general higher-dimension operators
suppressed by the flavor scale $M$.
The largest effects come from terms in the effective Lagrangian
of the form
\beq\eql{MFCNC}
\De\scr{L}_{\rm eff} \sim \myint d^2\th d^2\bar{\th}\,
\frac{1}{M^2} (P \bar{U})^\dagger (P \bar{U}),
\eeq
which give rise to mixing between the composite generations.
This translates to mixing masses between the composite generations of order
\beq
\frac{\De m^2_{jk}}{m_{\rm comp}^2} \sim \left(
\frac{\avg{\bar{U}}}{M} \right)^2 \sim y_{jk}.
\eeq
The most stringent bounds on squark mixing come from $K^0$--$\bar{K}^0$
mixing, and can be summarized as
\beq\eql{KKbar}
\Re \left( \frac{\De m^2_{\tilde{d}\tilde{s}}}{m_{\rm comp}^2} \right)
\lsim 10^{-1} \frac{m_{\rm comp}}{10\TeV},
\qquad
\Im \left( \frac{\De m^2_{\tilde{d}\tilde{s}}}{m_{\rm comp}^2} \right)
\lsim 10^{-2} \frac{m_{\rm comp}}{10\TeV}.
\eeq
Since $y_{ds} \sim 3 \times 10^{-4}$, this is easily satisfied even if we
assume that $C\!{}P$ violation in the flavor sector is maximal.

Note that the flavor sector generically introduces
additional contributions to the
third generation scalar masses from operators of the form
\beq\eql{addmass}
\De\scr{L} \sim \myint d^2\th d^2\bar{\th}\,
\frac{1}{M^2} \bar{U}^\dagger \bar{U} \Phi_3^\dagger \Phi_3,
\eeq
which gives
\beq
\De m_3^2 \sim \frac{N}{M^2} F_{\bar{U}}^2
\sim N \frac{\avg{\bar{U}}^2}{M^2} m_{\rm comp}^2.
\eeq
This can reasonably give contributions
to $m_3$ as large as $\sim 1\TeV$, large enough to cancel the
negative 2-loop contribution discussed above.

A striking signature of these models is that \emph{all} scalars of the
first two generations unify at the scale $\La_{\rm comp}$ (which need
not be close to the GUT scale).
The unification holds up to effects suppressed by a loop factor, and
is therefore expected to hold to $1\%$.
This striking pattern is difficult to obtain naturally in other models.

\subsection{`Dimensional Hierarchy' Models}
We next discuss `dimensional hierarchy' models that explain the observed
fermion mass hierarchy in terms of a hierarchy of dimensions of operators.
Specifically, we assume that the first-generation quarks and leptons correspond
to dimension 3 operators of the form $(P \bar{U} \bar{U})$,
second-generation quarks and leptons correspond to dimension 2 operators
$(P \bar{U})$, and third generation quarks and leptons are elementary
(dimension 1).
In this case, Yukawa couplings involving the composite states arise from
terms in the tree-level superpotential of the form
\beq\bal
\De W &= \frac{1}{M^4} H (P \bar{U} \bar{U}) (P \bar{U} \bar{U})
+ \frac{1}{M^3} H (P \bar{U} \bar{U}) (P \bar{U})
+ \frac{1}{M^2} H \Phi_3 (P \bar{U} \bar{U})
\\
&\qquad\qquad
+ \frac{1}{M^2} H (P \bar{U}) (P \bar{U})
+ \frac{1}{M} H \Phi_3 (P \bar{U}),
\eal\eeq
giving rise to a Yukawa matrix of the form
\beq
y \sim \pmatrix{
\ep^4 & \ep^3 & \ep^2 \cr
\ep^3 & \ep^2 & \ep   \cr
\ep^2 & \ep   & 1     \cr},
\qquad
\ep \sim \frac{\avg{\bar{U}}}{M}.
\eeq
This structure reproduces the main features of the observed fermion
mass hierarchy for $\ep \sim 10^{-1}$.

In this scenario there is no approximate flavor symmetry at low
energies because the first- and second-generation fields belong to different
strong-interaction multiplets.
We therefore have
\beq
\frac{\De m^2_{\tilde{d}\tilde{s}}}{m_{\rm comp}^2} \sim \sin\th_{\rm c}
\sim 10^{-1}.
\eeq
Comparing with the bounds from the $K^0$--$\bar{K}^0$ system \Eq{KKbar},
we see that for $m_{\rm comp} \sim 10\TeV$ we require either a $10\%$
fine-tuning or a $10\%$ suppression of $C\!{}P$-violating effects in the squark
masses.

A striking signature of these models is that the first- and second-generation
scalars unify in two multiplets at the scale $\La_{\rm comp}$.

As in the `meson' models considered above,
operators of the form \Eq{addmass} can give additional positive contributions
to the third-generation scalar masses in this class of models of order
$1\TeV$.

\section{Model Building}
\subsection{A Two Generation `Meson' Model}
We construct a model with two complete generations of quarks and leptons
corresponding to dimension-2 composite operators.
The model has $G_{\rm comp} = SU(15)$,
$G_{\rm lift} = SU(13)$, and is based on the `fundamentals only'
model analyzed in the Appendix.
The matter content is
\beq\nonumber
\begin{tabular}{c|cc|ccc}
 & $SU(13)$ & $SU(15)$  & $SU(15)$ & $SU(3)$ \\
\hline
$Q$ & \fund & \fund & {\bf 1} & {\bf 1} \\
$L$ &  $\overline{\fund}$  & {\bf 1}  & $\overline{\fund}$
& {\bf 1} \\
$\bar{U}$  & {\bf 1} & $\overline{\fund}$ & \fund & {\bf 1}\\
$\bar{D}$ & {\bf 1} & $\overline{\fund}$ & {\bf 1} & {\bf 1} \\
$S$ & {\bf 1} & \fund & {\bf 1} & \fund \\
\end{tabular}
\eeq
\noindent
The standard model gauge group $\SU{5}_{\rm SM}$ is embedded
in the fundamental of $SU(15)$ as%
\footnote{Here $SU(5)_{\rm SM}$ is only a shorthand for the standard
embedding of $SU(3)_{\rm C} \times SU(2)_{\rm W} \times U(1)_{\rm Y}$.}
\beq
\fund \to \bb{10} \oplus \mybar{\bb{5}}.
\eeq
The theory has the tree-level superpotential
\beq
W = \la Q L \bar{U} + m \bar{D} S.
\eeq
We assume that $\La_{15} \gg \La_{13}, m$.

The composite spectrum below the scale $\La_{15}$ is
\beq\nonumber
\begin{tabular}{cc|ccc}
Higgs & composite  & $SU(15)$ & $SU(3)$ \\
\hline
$\bar{D}$ & $\bar{D} \bar{U}^{N - 1}$
& $\overline{\fund}$ & {\bf 1} \\
$S$ & $S \bar{U} $ & \fund & \fund \\
\end{tabular}
\eeq
The operator corresponding to the composite states is given in both the
`Higgs' and `composite' description, along with their quantum numbers under
the global symmetries (see the Appendix for more details).
The mass term breaks the global symmetry $SU(3) \to SU(2)$,
and gives the `baryon' composites
and one of the `meson' composites
masses of order $m$.
It may appear unattractive to have an explicit mass term in the model,
but the low-energy behavior of the model is independent of the
value of $m$ as long as $100\GeV \lsim m \ll \La_{15}$.%
\footnote{Remarkably, the low-energy dynamics is insensitive to the relative
size of $m$ and $\La_{13}$; see the Appendix.}
In a more fundamental theory, the mass term may arise dynamically, or from
the VEV of a singlet field.

Below the scale $m$, the composite spectrum is therefore
\beq\nonumber
\begin{tabular}{cc|ccc}
Higgs & composite  & $SU(15)$ & $SU(2)$ \\
\hline
$S$ & $S \bar{U} $ & \fund & \fund \\
\end{tabular}
\eeq
The `horizontal' $SU(2)$ symmetry is broken by the Yukawa couplings
$\la$ and by
higher-dimension terms required to give fermion masses, but the
approximate $SU(2)$ symmetry is sufficient to suppress FCNC's,
as discussed in Section 2 above.
In this model, all the squarks and sleptons of the first two generations
unify together at the scale $\La_{15}$.

This model has 29 generations of `preons'
above the compositeness scale (plus Higgs fields),
and so the $SU(5)_{\rm SM}$ couplings have a
Landau pole within a decade of the compositeness scale.

This is still
compatible with perturbative unification if the compositeness scale
is \emph{above} the GUT scale.
The Landau pole may have a physical interpretation in terms of a `dual'
model.
(In fact, since the model contains only fundamentals of all the gauge group
factors, it is straightforward to construct a dual for any of the gauge
group factors individually.)
However, it is certainly unattractive that this model requires new
physics so close to the compositeness scale, and it is our hope that more
economical models with smaller matter content can be found.

\subsection{A Composite $\bar{\bf 5}$ `Meson' Model}
A simple way to avoid Landau poles near the compositeness scale using the
model-building technology developed in this paper is to have fewer states
composite.
For example, we can take $G_{\rm comp} = SU(5)$, $G_{\rm lift} = SU(3)$,
with the following matter content:
\beq\nonumber
\begin{tabular}{c|cc|ccc}
 & $SU(3)$ & $SU(5)$  & $SU(5)_{\rm SM}$ & $SU(3)$ \\
\hline
$Q$ & \fund & \fund & {\bf 1} & {\bf 1} \\
$L$ &  $\overline{\fund}$  & {\bf 1}  & $\overline{\fund}$
& {\bf 1} \\
$\bar{U}$  & {\bf 1} & $\fund$ & $\overline{\fund}$ & {\bf 1}\\
$\bar{D}$ & {\bf 1} & $\overline{\fund}$ & {\bf 1} & {\bf 1} \\
$S$ & {\bf 1} & \fund & {\bf 1} & \fund \\
\end{tabular}
\eeq
\noindent
with a tree-level superpotential
\beq
W = \la Q L \bar{U} + m \bar{D} S.
\eeq
We take $\La_5 \gg \La_3, m$.
The mass term breaks the global $SU(3)$ symmetry down to $SU(2)$,
and the light composite spectrum below the scale $m$
consists of two $\bar{\bf 5}$ of the $SU(5)_{\rm SM}$.

Above the compositeness scale $SU(5)_{\rm SM}$ has
$6 \times \bar{\bf 5} \oplus 3 \times {\bf 5}$ (rather than
$3 \times {\bf 5}$), and so there is no problem with a Landau
pole near the compositeness scale.

In this model, the masses of the scalars $\tilde{d}_L$, $\tilde{s}_L$,
$\tilde{e}_L$, and $\tilde{\nu}_L$ unify at the compositeness scale.
Of course, it is possible to make similar models where a small number
of states (such as the first two generation ${\bf 10}$'s) are composite,
and their masses will unify.

\subsection{An `Efficient' but Un-unified Model}
We now consider a model that gives two generations of composite quarks
without generating a Landau pole near the compositeness scale.
However, the composite states in this model do not arise in complete
$SU(5)$ multiplets, and so the model cannot be naturally embedded in a
grand-unified model.

The model is again based on the `fundamentals only'
model analyzed in the Appendix.
The field content is
\beq\nonumber
\begin{tabular}{c|cc|ccc}
 & $SU(2)$ & $SU(N)$  & $SU(N - 2)_L$ & $SU(2)$ & $SU(N - 2)_R$ \\
\hline
$Q$ & \fund & \fund & {\bf 1} & {\bf 1} & {\bf 1} \\
$L$ &  $\fund$  & {\bf 1}  & $\overline{\fund}$
& {\bf 1} & {\bf 1} \\
$L'$ &  $\fund$  & {\bf 1}  & {\bf 1}
& $\fund$ & {\bf 1} \\
$\bar{U}$  & {\bf 1} & $\overline{\fund}$ & $\fund$ & {\bf 1}  & {\bf 1} \\
$\bar{U}'$  & {\bf 1} & $\overline{\fund}$ & {\bf 1} & $\fund$  & {\bf 1} \\
$\bar{D}$ & {\bf 1} & $\overline{\fund}$ & {\bf 1} & {\bf 1} & {\bf 1} \\
$S$ & {\bf 1} & \fund & {\bf 1} & {\bf 1} & \fund \\
$S'$ & {\bf 1} & \fund & {\bf 1} & {\bf 1} & {\bf 1} \\
$X$ & {\bf 1} & {\bf 1} & {\bf 1} & $\fund$ & {\bf 1} \\
\end{tabular}
\eeq
with tree-level superpotential
\beq
W = \la Q L \bar{U} + \la' Q L' \bar{U}'
+ y X S \bar{U}'
+ m \bar{D} S'.
\eeq
The mass term gives mass of order $m$ to all composite states containing $S'$,
while the Yukawa coupling proportional to $y$ gives rise to a mass of order
$y \La_N$ to all composite states containing $\bar{U}'$.
The remaining composite states are
\beq\nonumber
\begin{tabular}{cc|ccc}
Higgs & composite  & $SU(N - 2)_L$ & $SU(N - 2)_R$ \\
\hline
$S$ & $S \bar{U} $ & \fund & \fund \\
\end{tabular}
\eeq
In the Higgs description, $S$ refers to only the $N - 2$ `colors'
that are orthogonal to $\bar{U}'$.

We can use this to construct a model with two composite generations of
quarks using the embedding of \Ref{lm}.
We take $N = 9$ and embed the standard-model
$\hbox{\rm gauge} \times \hbox{\rm flavor}$ group
\beq\eql{SMgroup}
\bal
SU(3)_C &\times SU(2)_W \times U(1)_Y
\times [ SU(2)_q \times SU(2)_u \times SU(2)_d \times U(1)_B ]
\\
&\subset SU(7)_L \times SU(7)_R
\eal
\eeq
where $SU(2)_{q,u,d}$ are flavor symmetry groups, and $U(1)_B$ is
baryon number (which has no anomalies under the strong groups).
The embedding is
\beq
SU(7)_L\ :\
\fund &\to ({\bf 1}, \fund)_0
\times (\fund, {\bf 1}, {\bf 1})_{\frac{1}{7}}
\nonumber\\
&\qquad \oplus (\overline{\fund}, {\bf 1})_{-\frac{1}{3}}
\times ({\bf 1}, {\bf 1}, {\bf 1})_{-\frac{4}{21}},
\\
SU(7)_R\ :\
\fund &\to ({\bf 1}, {\bf 1})_{-1}
\times ({\bf 1}, \overline{\fund}, {\bf 1})_{-\frac{1}{7}}
\nonumber\\
&\qquad \oplus ({\bf 1}, {\bf 1})_{1}
\times ({\bf 1}, {\bf 1}, \overline{\fund})_{-\frac{1}{7}}
\nonumber\\
&\qquad \oplus (\fund, {\bf 1})_{\frac{1}{3}}
\times ({\bf 1}, {\bf 1}, {\bf 1})_{\frac{4}{21}}.
\eeq
This gives rise to two composite generations of quarks, with additional
fields transforming under the group \Eq{SMgroup} as
\beq\bal
\Phi_u &\sim ({\bf 1}, \fund)_{-1}
\times (\fund, \overline{\fund}, {\bf 1})_0,
\\
\Phi_d &\sim ({\bf 1}, \fund)_{1}
\times (\fund, {\bf 1}, \overline{\fund})_0,
\\
A & \sim ({\bf 8}, {\bf 1})_0
\times ({\bf 1}, {\bf 1}, {\bf 1})_0,
\\
B &\sim ({\bf 1}, {\bf 1})_0
\times ({\bf 1}, {\bf 1}, {\bf 1})_0.
\eal\eeq
The fields $\Phi_{u,d}$ are `flavored' Higgs fields that may play a role
in flavor physics.
Alternatively, all of the extra fields above can be given masses of order
$\La_{9}$ by adding extra fields $X$ with conjugate quantum numbers and
adding Yukawa couplings of the form $\De W = X S \bar{U}$.
The composite {\bf 8} can be eliminated by a higher-dimension operator
of the form $\De W = (S \bar{U})^2$.

Above the composite scale, this model has 12 extra $SU(2)_W$ doublets,
and 5 extra $SU(3)_C$ flavors, so the Landau pole is not close to the
compositeness scale.

\subsection{A `Dimensional Hierarchy' Model}
We next consider a model in which the hierarchy of Yukawa couplings is
explained by the different dimensionalities of composite states in the
different generations.
The model we consider is based on the `antisymmetric tensor' model analyzed
in the Appendix.
The matter content is
\beq\nonumber
\begin{tabular}{c|cc|ccc}
 & $SU(2)$ & $SU(N)$ & $SU(N)$ & $SU(2)$ \\
\hline
$Q$ & \fund & \fund & {\bf 1}  & {\bf 1}    \\
$L$ & $\afund$ & {\bf 1} & $\afund$& {\bf 1}   \\
$\bar{U}$ & {\bf 1} & $\afund$ & \fund  & {\bf 1}     \\
$A$ & {\bf 1} & \asymm & {\bf 1}   & {\bf 1}   \\
$S$ & {\bf 1} & \fund & {\bf 1}  & \fund  \\
\end{tabular}
\eeq
where we have left $N$ arbitrary for the moment.
The tree-level superpotential is
\beq
W = \la Q L \bar{U}.
\eeq

As shown in the Appendix,
the composite spectrum of this model below the scale $\La_N$ is
\beq\nonumber
\begin{tabular}{cc|cc}
Higgs & composite  & $SU(N)$ \\
\hline
$A$ & $A \bar{U}^2$ & $\asymm\vphantom{\raisebox{3pt}{\asymm}}$ \\
$S$ & $S \bar{U} $ & \fund  \\
\end{tabular}
\eeq
We embed the standard model into the global $SU(N)$ by taking
$N = 15 + n$ with
\beq
\fund \to \bb{10} \oplus \mybar{\bb{5}} \oplus (n \times \bb{1}).
\eeq
Then the composite $\asymm$ state decomposes as
\beq
\asymm \to (n \times \bb{10}) \oplus (n \times \mybar{\bb{5}}) \oplus
\left[ {\bf 45} \oplus \mybar{{\bf 45}} \oplus
\mybar{{\bf 10}} \oplus \bb{5}
\oplus \left( \sfrac{1}{2} n(n - 1) \times {\bf 1} \right) \right].
\eeq
Allowing all possible superpotential terms among the $\asymm$ composites
gives mass to all of the states in brackets above, leaving $n - 1$
composite generations \cite{ALT}.
The largest mass term for the mirror family ($\mybar{{\bf 10}} \oplus \bb{5}$)
is with one linear combination of the dimension two families, since this is
the lowest dimension operator.
If we take $n = 1$ ($N = 16$), we obtain one composite generation from the
dimension-2 $\fund$ composite and one from the dimension-3 $\asymm$
composite.

As discussed in Section 2, there is no symmetry relating the
the first and second generation squarks, so the composite scalars
must have masses of order $10\TeV$ to suppress FCNC's.

This model has 18 generations of `preons' above the compositeness scale
(plus Higgs fields), and the Landau pole for the standard-model couplings
is approximately a factor of $10^2$ above the compositeness scale.

\subsection{An `Efficient' but Speculative Model}
To obtain more elegant models we would like find examples with smaller gauge
groups and matter content so that there are no Landau poles close to
the compositeness scale.
We now present a model with an efficient group-theoretic embedding,
but whose dynamics we do not know how to analyze completely.
If we make a reasonable dynamical assumption, this model gives rise to
compositeness and SUSY breaking by the mechanism discussed in Section 2.
The particle content is:
\beq\nonumber
\begin{tabular}{c|cc|cc}
 & $SU(k)$ & $SO(10)$ & $SU(10)$ & $SU(2)$  \\
\hline
$Q$ & $\fund$ & $\fund$ & {\bf 1} & {\bf 1}   \\
$L$ & $\afund$ & {\bf 1} & $\fund$ & {\bf 1}   \\
$\bar{U}$ & {\bf 1} & $\fund$ & $\afund$ & {\bf 1}   \\
$S$ & {\bf 1} & ${\bf 16}$ & {\bf 1} & $\fund$   \\
\end{tabular}
\eeq
with the usual tree-level superpotential
\beq
W = \la Q L \bar{U}.
\eeq

For $\avg{\bar{U}} \gg \La_{10}$,
$SO(10) \times SU(10)$ is broken to the diagonal $SO(10)$ subgroup and
$SU(k)$ gaugino condensation gives rise to a dynamical superpotential
\beq\eql{SOwdyn}
W_{\rm dyn} \sim \bar{U}^{10/k}.
\eeq
The potential therefore slopes toward $\bar{U} \to 0$ for $k < 10$.

The dynamics for small values of $\avg{\bar{U}}$ involves the strong-coupling
behavior of the $SO(10)$ gauge theory with spinors, which is presently
not well understood.
The $SO(10)$ gauge theory has a dual description in terms of an
$SU(2) \times SU(7+k)$ gauge theory with a complicated set of matter
representations, including a symmetric tensor of $SU(7+k)$ \cite{SO10duals}.
This dual is not weakly coupled in the infrared, so we cannot use it
to determine the behavior of the \Kahler potential for $\bar{U}$ in any
simple way.
Based on analogies with similar duals, one expects this theory to be at
a fixed point in the infrared \cite{SO10duals,fiveeasy}.

If we assume that the anomalous dimension of $\bar{U}$ is sufficiently
large,
then there is a local SUSY-breaking minimum with
$\avg{\bar{U}} \sim \La_{10}$.
The fermions from the {\bf 16}'s are exactly massless far from the origin, and
because there can be no phase transitions as a function of moduli they
are massless at the local minimum as well.
This model therefore contains two composite fermionic {\bf 16}'s which can
be identified with two standard-model generations (with right-handed neutrinos)
if we embed the standard model into $SU(10)$ via the standard GUT embedding
\beq
SU(10) \to SO(10)_{\rm SM}.
\eeq
FCNC's are suppressed by the approximate global $SU(2)$ symmetry of the
strong dynamics.
Above the compositeness scale $\La_{10}$, this model has
$3 + k/2$ additional `preonic' generations.
Note that for \eg  $k = 9$, only a $10\%$ change of the $\bar{U}$ scaling
dimension is required to obtain a local minimum, and the preonic theory
only has $7.5$ extra generations.
It is therefore very reasonable to assume that this model works and gives
a highly `efficient' composite model.

Yukawa couplings for the composite generations can be induced if we include
a Higgs field, $H$ embedded in the $\fund$ of the global $SU(10)$ by operators
of the form
\beq
\De W =
\frac{1}{M} S S H  \bar{U}
\eeq
This gives Yukawa couplings $y \sim \avg{\bar{U}} / M$ for the composite
quarks and leptons.
(Comparing to our previous expressions,
this corresponds to the composite operators
being dimension-$\sfrac{3}{2}$ operators.)
Thus the flavor scale $M$ can be pushed up even higher in
this model, and FCNC's are even more suppressed than in our
`meson' models.
Mixing between the composite generations and the fundamental third
generation $\Phi_3$ is more difficult to obtain.
It may arise from operators such as
\beq
\De W =
\frac{1}{M^2} \Phi_3 S S  H \bar{U}
\eeq
provided the right-handed sneutrino components of $S$ get VEVs.
Presumably these VEVs must occur below the scale  $\avg{\bar{U}}$.
If they are an order of magnitude
below this scale then we get an adequate suppression of  mixings with the
third generation.
It would be interesting to extend the model to include the
generation of neutrino masses through a seesaw with masses for the
composite right-handed neutrinos, but we will not pursue this subject here.

In the above discussion,
we have used a `Higgs' description where the $SU(10)$ gauge
dynamics is spontaneously broken.
This model has no `confined' description in the naive sense, since
we cannot write a composite chiral operator transforming as a {\bf 16} under
the unbroken $SO(10)$ global symmetry.
This model therefore does not exhibit `complementarity'  \cite{complement},
and a deeper
understanding of its strong-coupling behavior would
be very desirable.

%
%
%
%
%
%
%

\section{Conclusions}
We have shown that there is a wide class of realistic
models which dynamically break SUSY
and produce composite quarks and leptons,
all in a single strongly-coupled sector.
These models are remarkably simple.
For example, the complete model presented in Section 3.1
has particle content
\beq\nonumber
\begin{tabular}{c|cc|ccc}
 & $SU(13)$ & $SU(15)$  & $SU(15)$ & $SU(3)$ \\
\hline
$Q$ & \fund & \fund & {\bf 1} & {\bf 1} \\
$L$ &  $\overline{\fund}$  & {\bf 1}  & $\overline{\fund}$
& {\bf 1} \\
$\bar{U}$  & {\bf 1} & $\overline{\fund}$ & \fund & {\bf 1}\\
$\bar{D}$ & {\bf 1} & $\overline{\fund}$ & {\bf 1} & {\bf 1} \\
$S$ & {\bf 1} & \fund & {\bf 1} & \fund \\
\end{tabular}
\eeq
\noindent
where the first two groups are strong gauge groups, and the
remaining groups are global symmetries of the strong interactions.
The standard model gauge group $SU(5)_{\rm SM}$
is embedded in the $SU(15)$ symmetry via
$\fund \to \asymm \oplus \bar{\fund}$.
The theory has a tree-level superpotential with the most
general couplings of the form
\beq
W = \la Q L \bar{U} + m \bar{D} S
+ \frac{1}{M^2} H (S\bar{U})^2
+ \frac{1}{M} H \Phi S \bar{U}
+ H \Phi^2
\eeq
where $H$ is a Higgs field and
$\Phi$ denotes an elementary (third-generation) quark or lepton
field.
This model gives rise to two full generations of composite quarks and
leptons;
breaks SUSY;
and gives rise to Yukawa couplings of the hierarchical form
\beq
y \sim \pmatrix{\ep^2 & \ep^2 & \ep \cr
\ep^2 & \ep^2 & \ep \cr
\ep & \ep & 1 \cr},
\qquad
\ep \sim \frac{\La_{15}}{M}.
\eeq
We see that many of the fermion mass hierarchies are automatic
consequences of approximate symmetries in this model.
Even if we assume that there is no GIM mechanism in the higher-dimension
operators that give rise to the Yukawa couplings, approximate flavor
symmetries of the strong interactions guarantee the natural suppression
of flavor-changing neutral currents (including $\ep_K$) with no fine-tuning.
The model is compatible with perturbative unification of gauge couplings
if the compositeness scale is at or above the GUT scale.

We have presented other models that are similarly simple, including a
model that generates hierarchical Yukawa couplings of the form
\beq
y \sim \pmatrix{\ep^4 & \ep^3 & \ep^2 \cr
\ep^3 & \ep^2 & \ep \cr
\ep^2 & \ep & 1 \cr},
\qquad
\ep \sim \frac{\La_{16}}{M}.
\eeq
All of these models have a very distinctive phenomenology:
the composite sfermions of the first two generation are heavier than the
gauginos, Higgsinos, and third-generation sfermions;
and the composite sfermions unify at the compositeness scale
(given by $\La_{15}$ in the first model and $\La_{16}$ in the second).

The main unattractive feature of these models is the large number of
`preon' fields charged under the standard model group, resulting in
a Landau pole for the standard-model interactions.
This can be avoided with the model-building technology presented here
in less ambitious models with fewer composite states.
However, another result of this paper is that the dynamics that gives
rise to simultaneous compositeness and SUSY breaking occurs in a large
class of models.
This includes models whose low-energy dynamics is governed by
confinement, non-trivial infrared fixed points, or free-magnetic
phases.
We believe that it is quite likely that further progress in understanding
the dynamics of SUSY gauge theories will lead to the discovery of many
additional models that display the dynamics illustrated here.

As an example, we presented in Section 3.5 a model that gives rise to
two composite generations, without Landau poles near the compositeness
scale.
This model cannot be completely analyzed, but it works in complete analogy
with models that we can analyze provided that a plausible inequality on
an anomalous dimension is satisfied.

Our final conclusion is that  the connection between
compositeness and SUSY breaking is worth further exploration
on both the theoretical and phenomenological fronts.

%

\section{Acknowledgments}
M.A.L. and J.T. thank the CERN theory group for hospitality
during part of this work, and thank N. Arkani-Hamed,
C. Cs\'aki, R. Rattazzi, H. Murayama, A. Nelson, and M. Schmaltz
for discussions.
M.A.L. is supported by a fellowship from the Alfred P. Sloan Foundation.
J.T. is supported by the National Science
Foundation under grant PHY-95-14797, and is also partially supported by
the Department of Energy under contract DE-AC03-76SF00098.

\appendix{Appendix A: Analysis of Models}
In this Appendix, we present the detailed analysis of gauge theories
of the type described in the main text.

\subsection{$SU \times SU$ Fundamentals Only}
This model has gauge group $SU(k) \times SU(N)$,
and matter content given by
\beq\nonumber
\begin{tabular}{c|cc|ccc}
 & $SU(k)$ & $SU(N)$  & $SU(N)$ & $SU(F)$ & $SU(N - k + F)$ \\
\hline
$Q$ & \fund & \fund & {\bf 1} & {\bf 1} & {\bf 1} \\
$L$ &  $\overline{\fund}$  & {\bf 1}  & $\overline{\fund}$
& {\bf 1} & {\bf 1} \\
$\bar{U}$  & {\bf 1} & $\overline{\fund}$ & \fund & {\bf 1} & {\bf 1}\\
$\bar{D}$ & {\bf 1} & $\overline{\fund}$ & {\bf 1} & \fund & {\bf 1} \\
$S$ & {\bf 1} & \fund & {\bf 1} & {\bf 1} & \fund \\
\end{tabular}
\eeq

\noindent
In addition, the theory has the usual tree-level superpotential
\beq
W = \la Q L \bar{U}.
\eeq
This model is interesting because it shows that s-confinement
\cite{sconfine} is
not necessary for our mechanism to work.
The result of the analysis is that for
\beq
F < k < N,
\qquad
F > 0,
\eeq
this model has a local SUSY-breaking minimum with
$\avg{\det\bar{U}} \sim (\La_N)^N$.
(For $F = 0$, the theory has a deformed moduli space and
we cannot determine whether there is a local
SUSY-breaking minimum.)
At the local minimum, the theory has composite states given by
\beq\nonumber
\begin{tabular}{cc|ccc}
Higgs & composite  & $SU(N)$ & $SU(F)$ & $SU(N - k + F)$ \\
\hline
$\bar{D}$ & $\bar{D} \bar{U}^{N - 1}$
& $\overline{\fund}$ & \fund & {\bf 1} \\
$S$ & $S \bar{U} $ & \fund & {\bf 1} & \fund \\
\end{tabular}
\eeq
The description of the composite states is given in both the `Higgs'
and `composite' description, along with their quantum numbers under
the unbroken global symmetries.


We now give the details of the analysis of this model.
The classical moduli space can be labeled by the gauge invariant
operators
\beq\nonumber
\begin{tabular}{c|ccc}
& $SU(N)$ & $SU(F)$ & $SU(N - k + F)$ \\
\hline
$\bar{U} S$ & \fund & {\bf 1} & \fund \\
$\bar{D} S$ & {\bf 1} & \fund & \fund \\
$L^k$ & $\overline{[k]}$ & {\bf 1} & {\bf 1} \\
$Q^k S^{N - k}$ & {\bf 1} & {\bf 1} & $\overline{[F]}$ \\
$\bar{U}^{N}$ & {\bf 1} & {\bf 1} & {\bf 1} \\
$\bar{U}^{N - 1} \bar{D}$ & $\afund$ & \fund & {\bf 1} \\
$\bar{U}^{N - 2} \bar{D}^2$ & $\overline{\asymm}$ & $\asymm$ & {\bf 1} \\
\vdots & \vdots & \vdots & \vdots \\
$\bar{U}^{N - F} \bar{D}^F$ & $\overline{[F]}$ & $[F]$ & {\bf 1} \\
\end{tabular}
\eeq
Here $[n]$ denotes the $n$-index antisymmetric tensor.
The moduli space has three branches, depending on which baryon
operators are nonzero:
\beq\bal
\hbox{$L$ branch:} & \quad L^k \ne 0 \\
\hbox{$Q$ branch:} & \quad Q^k S^{N - k} \ne 0 \\
\hbox{$\bar{U}$ branch:} & \quad
\bar{U}^{N}, \bar{U}^{N - 1} \bar{D}, \ldots,
\bar{U}^{N - F} \bar{D}^F \ne 0.
\eal\eeq
(This means that \eg\ if $L^k \ne 0$, all other baryon operators
vanish.)
We will be interested in the $\bar{U}$ branch.
There are classical constraints on the operators on this
branch that we will not write.

We assume $\La_N \gg \La_k$, and take arbitrary
VEV's on the $\bar{U}$ branch of the moduli space:
\beq
\avg{\bar{U}} &= \pmatrix{a_1 & & \cr & \ddots & \cr & & a_N \cr}.
\eeq
First consider the case when $a_1, \ldots, a_N \gg \La_N$,
then all components of $Q$ and $L$ get
massive, and $SU(k)$ gaugino condensation gives rise to a dynamical
superpotential that pushes $\det\bar{U}$ to zero:
\beq
W_{\rm dyn} \propto (\det\bar{U})^{1/k} \sim \bar{U}^{N/k}.
\eeq
We next consider the case
\beq
\eql{largeVEVs}
\La_k \ll a_1, \ldots, a_n &\ll \La_N,
\qquad
a_{n + 1}, \ldots, a_N \gsim \La_N.
\eeq
By taking all VEV's large compared to $\La_k$, we ensure that
the non-perturbative superpotential generated by the $SU(k)$
dynamics is less important than the $SU(N)$ $D$-term potential,
and therefore it makes sense to restrict attention to the
$\bar{U}$ branch of the classical moduli space.
We will now show that the small VEV's $a_1, \ldots, a_n$ are
driven to larger values by the non-perturbative $SU(k)$
dynamics.
We do this by constructing the effective theory below the
large VEV's $a_{n + 1}, \ldots, a_N$ classically and then
analyzing the non-perturbative dynamics in the resulting
low-energy theory.
In general, other VEV's must be large in order to solve the
classical $D$-flat constraints, and we must analyze \emph{all}
possibilities and show that $\det\bar{U}$ is driven away from
zero in all cases.


The analysis divides into two cases, depending on the VEV's for
$\bar{D}$.
First, suppose that the $D$-flat conditions for the large VEV's in
$\bar{U}$ are satisfied by having $n$ large VEV's for $D$, so
that the baryon operator $\bar{U}^{N - n} \bar{D}^{n}$ is large.
This requires $n \le F$.
In this case, the $SU(N)$ gauge group
is completely broken, and the only strong
dynamics is in the unbroken $SU(k)$ gauge group.
The fields $\bar{D}$ and $S$ may have additional large VEV's, but
these do not affect the analysis.
The large VEV's in $\bar{U}$ leave $n$ flavors of $SU(k)$ massless,
denoted by $q$, $\ell$.
The theory has a superpotential
\beq
W_{\rm eff} = \la q \ell \bar{u},
\eeq
where $\bar{u}$ contains the excitations of the small VEV's of
$\bar{U}$. Gaugino condensation for $SU(k)$ leads to a dynamical
superpotential
\beq\eql{smallwdyn}
W_{\rm dyn} \propto \bar{u}^{n/k}
\eeq
Thus the non-perturbative $SU(k)$ dynamics pushes $\bar{u}$ away from
the origin if $k > F$, since $n \le F < k$.

The remaining case is that the $D$-flat conditions for the large
VEV's in $\bar{U}$ are satisfied by having (at least)
$N - n$ components of $S$ large, so that the meson operator
$\bar{U} S$ is large.
This requires $n \ge k - F$.
(We need not consider the possibility that some components of $Q$ are
large, since this corresponds to a different branch of the moduli space.)
We first integrate out the fields that are massive due to the $N - n$
large $S$ VEV's to obtain a theory with gauge group
$SU(n) \times SU(k)$, and matter content given by
\beq\nonumber
\begin{tabular}{c|cc|ccc}
 & $SU(k)$ & $SU(n)$  & $SU(n)$ & $SU(F)$ & $SU(n - k + F)$ \\
\hline
$q$ & \fund & \fund & {\bf 1} & {\bf 1} & {\bf 1} \\
$\ell$ &  $\overline{\fund}$  & {\bf 1}  & $\overline{\fund}$
& {\bf 1} & {\bf 1} \\
$\bar{u}$  & {\bf 1} & $\overline{\fund}$ & \fund & {\bf 1} & {\bf 1}\\
$\bar{d}$ & {\bf 1} & $\overline{\fund}$ & {\bf 1} & \fund & {\bf 1} \\
$s$ & {\bf 1} & \fund & {\bf 1} & {\bf 1} & \fund \\
\end{tabular}
\eeq
with some singlets not shown.
The superpotential is
\beq\eql{wsmallagain}
W_{\rm eff} = \la q \ell \bar{u}.
\eeq
This is just the original model with $N$ replaced by $n$.
Some of the components of $\bar{d}$ and $s$ may also be large,
and this can be analyzed in the effective theory above.%
\footnote{A strict application of effective field theory ideology would
require us to integrate out all states with large mass at the same time.
However, the result is the same because the heavy modes can be
integrated out at tree level, ignoring the breaking of SUSY.}

Consider first the case where all the components of $\bar{d}$ and $s$
have small VEV's.
When $F \ge 2n$ the effective theory is infrared-free, so we see that
gaugino condensation for $SU(k)$ leads to a dynamical
superpotential
\beq
W_{\rm dyn} \propto \bar{u}^{n/k}
\eeq
which forces $\bar{u}$ to run away for $k > n$ which is always
satisfied provided that $k > \sfrac{1}{2} F$ since $F \ge 2 n$.

For $2 \le F \le 2n$, the $SU(N)$ dynamics has a dual description,
and the effective theory has gauge group $SU(k) \times SU(F)$,
with matter content given by
\beq\nonumber
\begin{tabular}{c|cc|ccc}
 & $SU(k)$ & $SU(F)$ & $SU(n)$ & $SU(F)$ & $SU(n - k + F)$ \\
\hline
$(\bar{d} q)$ & \fund & {\bf 1} & {\bf 1} & \fund & {\bf 1} \\
$(\bar{u} s)$ & {\bf 1} & {\bf 1} & \fund & {\bf 1} & \fund \\
$(\bar{d} s)$ & {\bf 1} & {\bf 1} & {\bf 1} & \fund & \fund \\
$\tilde{q}$ & $\afund$ & $\afund$ & {\bf 1} & {\bf 1}
& {\bf 1} \\
$\tilde{\bar{u}}$ & {\bf 1} & \fund & $\afund$ & {\bf 1} & {\bf 1} \\
$\,\tilde{\!\bar{d}\,}\!$ & {\bf 1} & \fund & {\bf 1} & $\afund$ & {\bf 1} \\
$\tilde{s}$ & {\bf 1} & $\afund$ & {\bf 1} & {\bf 1} & $\afund$ \\
\end{tabular}
\eeq
The theory has a dynamical superpotential
\beq\eql{Wprime}
W_{\rm dyn} = \frac{1}{\La_n} \left[
(\bar{d} q) \tilde{q} \,\tilde{\!\bar{d}\,}\!
+ (\bar{u} s) \tilde{s} \tilde{\bar{u}}
+ (\bar{d} s) \tilde{s} \,\tilde{\!\bar{d}\,} \right].
\eeq
In this theory the $SU(F)$ gauge dynamics is infrared free provided
$F \le \sfrac{1}{2} n$.
In this case, the non-perturbative $SU(k)$ dynamics dominates,
generating a dynamical superpotential
\beq
W_{\rm dyn} \sim \tilde{\!\bar{d}\,}{}^{F/k}
\eeq
which pushes $\,\tilde{\!\bar{d}\,}\!$ away from the origin for $k>F$.
Since the duality operator mapping is
$\bar{u}^N \leftrightarrow \tilde{\!\bar{d}\,}{}^F$, this means that
$\bar{u}$ is forced away from the origin.

The condition $F \le \sfrac{1}{2} n$ arose from demanding that the dual
theory is infrared free, but this is actually too restrictive
since the field $\bar{u}$ can be pushed away from the origin even
if the dual theory has an infrared fixed point (see Section \ref{SBC}).
In the conformal window for the dual ($\sfrac{1}{2} F < n < 2F$)
the scaling dimension of the operator $\tilde{\!\bar{d}\,}{}^F$ is
$\sfrac{3}{2} n F / (n + F)$.
The field $\bar{u}$ is therefore pushed away from the origin for
$k > \sfrac{3}{2} n F / (n + F)$.
(As $n \to 2F$, the dual becomes infrared free and we recover our
previous condition $k > F$, while when $n \to \sfrac{1}{2} F$, the
`electric' description becomes infrared free and we recover our
previous condition $k > \sfrac{1}{2} F$.)
Since this must be satisfied for all $k - F \le n \le N$, we have
\beq\eql{confbound}
k > \frac{3 N F}{2(N + F)}.
\label{conformal}
\eeq
Note that as we approach the end of the conformal window
($N \to \sfrac{1}{2} F$) this requires $k \to N$, which is
the marginal case of inverted hierarchy models.
In the conformal window, the bound (\ref{conformal})
is superseded\footnote{The bound
\Eq{confbound} will be important in the
next section where the baryons are removed from
the low energy theory.} by the result
$k>F$ obtained by studying the case with large VEV's for
the baryons $\bar{U}^{N-n} D^n$.

We have been considering the case where there are no additional large
VEV's for $\bar{d}$ and $s$ in the effective theory given above
\Eq{Wprime}.
These VEV's reduce the number of colors and flavors of the $SU(n)$
group in the effective theory by the same amount. For $F \ge 2n$ this gives
the effective theory a larger positive $\beta$ function, so the theory is
still infrared-free, and the same analysis applies. For $2 \le F \le 2n$,
the $SU(n)$ dynamics again has a dual description in terms of
an $SU(F)$ gauge group, but now with fewer flavors.
It is easily checked that the analysis above still applies and that
the $SU(k)$ gauge dynamics forces $\bar{u}$ away from
the origin for sufficiently large $k$.

Now go back to the effective theory described above \Eq{Wprime}
for the case $F = 1$, assuming that the fields $s$ and $\bar{d}$
do not have large VEV's.
The $SU(N)$ theory is then s-confining, and
the low-energy dynamics can be described by a theory with
$SU(k)$ gauge group and matter content given by
\beq\nonumber
\begin{tabular}{c|c|cc}
 & $SU(k)$ & $SU(n)$ & $SU(n - k + 1)$ \\
\hline
$(\bar{d} q)$ & \fund & {\bf 1} & {\bf 1} \\
$(\bar{u} s)$ & {\bf 1} & \fund & \fund \\
$(\bar{d} s)$ & {\bf 1} & {\bf 1} & \fund\\
$(\bar{u}^n)$ & {\bf 1} & {\bf 1} & {\bf 1} \\
$(\bar{u}^{n - 1} \bar{d})$ & {\bf 1} & $\overline{\fund}$ & \fund\\
$(q^{k} s^{n - k})$ & {\bf 1} & {\bf 1} & $\overline{\fund}$\\
$(q^{k - 1} s^{n - k + 1})$ & $\overline{\fund}$ & {\bf 1} & {\bf 1}\\
\end{tabular}
\eeq
The fields are indicated by composite operators with the
quantum numbers in parentheses.
The theory has a dynamical superpotential
\beq\eql{ssmallw}
\bal
W_{\rm dyn} &= \frac{1}{\La_n^{2n - 1}} \left[
(\bar{u}^n) (\bar{d} q) (q^{k - 1} s^{n - k + 1})
+ (\bar{u}^n) (\bar{d} s) (q^{k} s^{n - k}) \right.
\\
&\qquad \left. +\,
(\bar{u}^{n - 1} \bar{d}) (\bar{u} s) (q^{k} s^{n - k}) \right].
\eal\eeq
The $SU(k)$ gauge group has one flavor with a trilinear coupling
that pushes $(\bar{u}^n)$ away from the origin.
We must now consider the possibility that $\bar{d}$
and/or $s$ have large VEV's.
As in the dual case analyzed above, these VEV's change the number
of the $SU(n)$ colors and flavors by the same amount, and therefore
again lead to s-confinement \cite{sconfine}, and
the $SU(k)$ dynamics then pushes $\bar{u}$ away from the origin.

Finally, we consider the case $F = 0$, where
the $SU(n)$ theory has a deformed moduli space and the argument for a local
SUSY breaking minimum fails.
In this case, the effective theory has gauge group $SU(k)$
with matter content given by
\beq\nonumber
\begin{tabular}{c|c|cc}
 & $SU(k)$ & $SU(n)$ & $SU(n - k)$ \\
\hline
$(\bar{u} s)$ & {\bf 1} & \fund & \fund \\
$(\bar{u}^n)$ & {\bf 1} & {\bf 1} & {\bf 1} \\
$(q^{k} s^{n - k})$ & {\bf 1} & {\bf 1} & $\overline{\fund}$\\
\end{tabular}
\eeq
with quantum constraint
\beq
(\bar{u}^n) (q^{k} s^{n - k}) = \La_n^{2n}.
\eeq
The $SU(k)$ dynamics gives rise to a dynamical superpotential
\beq
W_{\rm dyn} \sim (\bar{u}^n)^{1/k},
\eeq
on this moduli space.
The criterion for a supersymmetric vacua is that the gradient of
$W_{\rm dyn}$ is proportional to the gradient of the constraint:
\beq\bal
\frac{\partial W_{\rm dyn}}{\partial (\bar{u}^n)}
= \frac{1}{k} (\bar{u}^n)^{1/k - 1} &\propto (q^{k} s^{n - k}),
\\
\frac{\partial W_{\rm dyn}}{\partial (q^{k} s^{n - k})}
= 0 &\propto (\bar{u}^n).
\eal\eeq
There are solutions $(\bar{u}^n) \to 0$, $(q^{k} s^{n - k}) \to \infty$
as well as $(\bar{u}^n) \to \infty$,
$(q^{k} s^{n - k}) \to 0$ (where the constant of proportionality vanishes).
However, we cannot control the K\"ahler potential sufficiently
in this case to know whether $\bar{u}$ is pushed away from the origin.

\subsection{$SU \times SU$ without `Baryon' Composites}
The existence of composite fermions with the quantum numbers of
high-dimension `baryon' operators in the previous model
is inconvenient for the type of
model-building we are interested in.
A simple way to eliminate the unwanted composite states in this model
is to add a mass term for all $\bar{D}$ fields to the
tree-level superpotential:
\beq
W = \la Q L \bar{U} + m \bar{D} S.
\eeq
This breaks the global symmetry $SU(N - k + F) \to SU(N - k)$.
For $m \gg \La_N$, we can analyze the non-perturbative dynamics by
first integrating out $\bar{D}$ and $S$.
We then obtain the $F = 0$ model, for which we cannot establish the
existence of a SUSY-breaking minimum.
However, for $m \ll \La_N$, we will show that the `baryon' composite
fields acquire a mass of order $m$, and the composite spectrum below
the scale $m$ is as follows:
\beq\nonumber
\begin{tabular}{cc|cc}
Higgs & composite  & $SU(N)$ & $SU(N - k)$ \\
\hline
$S$ & $S \bar{U} $ & \fund & \fund \\
\end{tabular}
\eeq
Remarkably, this result does not depend on the relative sizes of $m$
and $\La_k$.
We find that for
\beq\eql{range}
\bal
F < k < N,
\qquad
0 < F \le \frac{N}{2}, \\
 \frac{3 N F}{2(N + F)} < k < N,
\qquad
\frac{N}{2} < F \le 2N,
\eal\eeq
this model has a local SUSY-breaking minimum.

We first consider the possibility that $\bar{D}$ has large VEV's
(in the sense of \Eq{largeVEVs}).
In this case, the $SU(N)$ gauge symmetry is completely broken, and
the $SU(k)$ dynamics generates a dynamical superpotential that is
independent of $\bar{D}$ (see \Eq{smallwdyn}).
Therefore, the potential for for $\bar{U}$ slopes away from the origin,
while the potential for $\bar{D}$ slopes toward zero.%
%

The remaining cases have $N - n$ components of $S$ large.
The effective theory is given above \Eq{wsmallagain}, with the
superpotential modified to
\beq
W_{\rm eff} = \la q \ell \bar{u} + m \bar{d} s.
\eeq

We now analyze the effective theory for various values of $F$.
For $F \ge 2n$ the effective theory is infrared-free, so we see that
gaugino condensation for $SU(k)$ leads to a
superpotential
\beq
W_{\rm dyn} \propto \bar{u}^{n/k} + m \bar{d} s
\eeq
which forces $\bar{u}$ to run away for $k > n$ which is always
satisfied provided that $k > \sfrac{1}{2} F$ since $F \ge 2n$.

For $2 \le F \le 2n$ the theory has a dual description given near
\Eq{Wprime}, with the dynamical superpotential modified to
\beq
W_{\rm dyn} = \frac{1}{\La_n} \left[
(\bar{d} q) \tilde{q} \,\tilde{\!\bar{d}\,}\!
+ (\bar{u} s) \tilde{s} \tilde{\bar{u}}
+ (\bar{d} s) \tilde{s} \,\tilde{\!\bar{d}\,} \right]
+ m (\bar{d} s).
\eeq
The presence of the linear term in $(\bar{d} s)$ does not prevent
the theory from forcing $\,\tilde{\!\bar{d}\,}\!$ away from the
origin.
Consider giving a VEV for $\,\tilde{\!\bar{d}\,}\!$ satisfying
\beq
\La_k, (m \La_n)^{1/2} \ll \,\tilde{\!\bar{d}\,}\! \ll \La_N.
\eeq
In this case, the fields $(\bar{d}q)$, $\tilde{q}$ and
$(\bar{d} s)$, $\tilde{s}$ get masses of order
$\,\tilde{\!\bar{d}\,}\!$.
The linear term implies
$\tilde{s} \sim \La_n m / \,\tilde{\!\bar{d}\,}\!
\ll \,\tilde{\!\bar{d}\,}\!$, which can be treated as a small
perturbation.
Below the scale $\,\tilde{\!\bar{d}\,}\!$,
the low-energy effective theory is again $SU(k)$ super Yang--Mills
theory.
$SU(k)$ gaugino condensation then gives rise to a dynamical superpotential
that pushes $\,\tilde{\!\bar{d}\,}\!$ away from the origin as before.

Note that for $\,\tilde{\!\bar{d}\,}\! \sim \La_n$, the $(\bar{d}s)$
equation of motion implies that $\tilde{s} \sim m$, which gives a mass
of order $m$ to $\tilde{\bar{u}}$.
This is how the `baryon' composites get mass in the dual description.

For $F = 1$, the story is very similar.
In this case, the theory s-confines \cite{sconfine}, and the effective
theory is given near \Eq{ssmallw},
with the dynamical superpotential modified to
\beq
\bal
W_{\rm dyn} &= \frac{1}{\La_n^{2n - 1}} \left[
(\bar{u}^n) (\bar{d} q) (q^{k - 1} s^{n - k + 1})
+ (\bar{u}^n) (\bar{d} s) (q^{k} s^{n - k}) \right.
\\
&\qquad \left. +\,
(\bar{u}^{n - 1} \bar{d}) (\bar{u} s) (q^{k} s^{n - k}) \right]
+ m (\bar{d} s).
\eal\eeq
Give a VEV to $(\bar{u}^n)$ satisfying
\beq
\La_k^n, (\La_n^{2n - 1} m)^{1/2} \ll (u^n)
\ll \La_n^n.
\eeq
This VEV gives $(\bar{d} q), (q^{k - 1} s^{n - k + 1})$
and $(\bar{d} s), (q^{k} s^{n - k})$ a mass of order $\bar{u}$;
the presence of the linear term gives the field
$(q^{k} s^{n - k})$ a small VEV.
Below the scale $\bar{u}$, the effective theory is again $SU(k)$
super Yang--Mills, and the dynamical superpotential pushes
$\bar{u}$ away from the origin.

For $(\bar{u}^n) \sim \La_n^n$, the $(\bar{d}s)$ equation of motion
implies that $(\bar{u}^{n - 1} \bar{d})$ gets a mass of order $m$.
This is how the `baryon' composites get mass in the confined description.

Putting together the inequalities for these various cases to work, we
arrive at \Eq{range}.
Note that the fact that the VEV $\bar{D}$ is always
pushed toward the origin allows this model to work for a wider range of
parameters than the massless model considered above.

\subsection{Antisymmetric Tensors}
We now turn to a model with dimension-2 and dimension-3 composites.
\beq\nonumber
\begin{tabular}{c|cc|ccc}
 & $SU(k)$ & $SU(N)$ & $SU(N)$ & $SU(F - k)$ & $SU(F-4)$\\
\hline
$Q$ & \fund & \fund & {\bf 1} & {\bf 1}  & {\bf 1} \\
$L$ & $\afund$ & {\bf 1} & $\afund$ & {\bf 1} & {\bf 1}  \\
$\bar{U}$ & {\bf 1} & $\afund$ & \fund & {\bf 1} & {\bf 1}  \\
$\bar{D}$ & {\bf 1} & $\afund$  & {\bf 1}  & {\bf 1}  & \fund \\
$A$ & {\bf 1} & \asymm & {\bf 1} & {\bf 1} & {\bf 1}  \\
$S$ & {\bf 1} & \fund & {\bf 1} & \fund & {\bf 1}  \\
\end{tabular}
\eeq
The model has the usual superpotential
\beq
W = \la Q L \bar{U}.
\eeq
For $F= k = 4$, this is the model analyzed in \Ref{ALT}.
The composite fermion spectrum is
\beq\nonumber
\begin{tabular}{cc|ccc}
Higgs & composite  & $SU(N)$ & $SU(F - k)$& $SU(F-4)$ \\
\hline
$\bar{D}$ & $\bar{D} \bar{U}^{N-1} $ & $\afund$ & {\bf 1} & \fund  \\
$A$ & $A \bar{U}^2$ & $\asymm\vphantom{\raisebox{3pt}{\asymm}}$ & {\bf 1} &
{\bf 1} \\
$S$ & $S \bar{U} $ & \fund & \fund & {\bf 1} \\
\end{tabular}
\eeq

We now briefly analyze this model.
For $\det\bar{U} \gg (\La_N)^N$, $SU(k)$
gaugino condensation gives rise to (an exact) dynamical superpotential
\beq
W \propto (\det\bar{U})^{1/k} \sim \bar{U}^{N/k}.
\eeq

For $F=4$ the $SU(N)$ interactions s-confine \cite{sconfine,Pouliot},
and the analysis follows that
of \Ref{ALT}.   For odd $N$ ($N = 2n+1$) and $k=2$
the moduli space is parameterized by
\beq\nonumber
\begin{tabular}{c|cc}
 & $SU(N)$   & $SU(2)$ \\
\hline
$S \bar{U}$ & \fund &  \fund
\\
$S A^n$  & \hbox{\bf 1} &   \fund
\\
$S Q^2 A^{n - 1}$  & \hbox{\bf 1} &  \fund
\\
$A \bar{U}^2$ & \asymm &  {\bf 1}
\\
$\bar{U}^{N}$ & \hbox{\bf 1} &  {\bf 1}
\\
$L^2$  & $\afund$ &  {\bf 1}
\end{tabular}
\eeq

The degrees of freedom below the scale $\La_N$ are
\beq\nonumber
\begin{tabular}{c|c|cc}
 & $SU(2)$  & $SU(N)$ & $SU(2)$ \\
\hline
$(Q\bar{U})$ & \fund & \fund & {\bf 1}
\\
$(S \bar{U})$ & {\bf 1} & \fund &  \fund
\\
$(Q A^n)$ & \fund & \hbox{\bf 1} &  {\bf 1}
\\
$(S A^n)$ & {\bf 1} & \hbox{\bf 1} &   \fund
\\
$(S Q^2 A^{n - 1})$ & {\bf 1} & \hbox{\bf 1} &  \fund
\\
$(S^2 Q A^{n - 1})$ & \fund & \hbox{\bf 1} &   {\bf 1}
\\
$(A \bar{U}^2)$ & \hbox{\bf 1} & \asymm &  {\bf 1}
\\
$(\bar{U}^{N})$ & \hbox{\bf 1} & \hbox{\bf 1} &  {\bf 1}
\\
$L$ & $\fund$ & $\afund$ &  {\bf 1}
\end{tabular}
\eeq
The effective superpotential is given by the sum of the tree
superpotential and a dynamical superpotential \cite{sconfine,Pouliot}:
\beq\bal
W_{\rm eff} &=
 {{1}\over{\Lambda_N^{2N-1}}}
\biggl\{
\left[ (Q A^n) (Q \bar{U})^3 (S \bar{U})^2
+(S A^n) (S \bar{U})  (Q \bar{U})^2 \right] (A \bar{U}^2)^{n - 1} \\
&\qquad\qquad\quad
+ (S Q^2 A^{n - 1}) (S \bar{U}) (A \bar{U}^2)^n
+ (S^2 Q A^{n - 1}) (Q\bar{U}) (A \bar{U}^2)^n
\\
&\qquad\qquad\quad
+ (\bar{U}^{N}) (S A^n) (S Q^2 A^{n - 1})
+ (\bar{U}^{N}) (Q A^n) (S^2 Q A^{n - 1})\biggr\}
\\
&\qquad
+ \la L (Q\bar{U}).
\eal\eeq
Integrating out $L$ and $(Q \bar{U})$ leaves $SU(2)$ with one flavor
$(Q A^n)$ and $(S^2 Q A^{n-1})$ as well as some singlets with a
superpotential
\beq\bal
W_{\rm eff} &=
{{1}\over{\Lambda_N^{2N-1}}}
\biggl[
 (S Q^2 A^{n - 1}) (S \bar{U}) (A \bar{U}^2)^n
+ (\bar{U}^{N}) (S A^n) (S Q^2 A^{n - 1})
\\
&\qquad\qquad\quad
+ (\bar{U}^{N}) (Q A^n) (S^2 Q A^{n - 1})\biggr].
\eal\eeq
The last two terms in this superpotential are mass terms when the baryon
$(\bar{U}^N)$
has a VEV, so on this branch of the moduli space we find a dynamical
superpotential:
\beq
W_{\rm dyn} \sim (\bar{U}^N)^{1/2},
\eeq
which forces the baryon to run away for any $N$.

For odd $N$ ($N = 2n+1$) and $k=3$
The moduli space is parameterized by
\beq\nonumber
\begin{tabular}{c|c}
  & $SU(N)$  \\
\hline
$S \bar{U}$  & \fund
\\
$S A^n$  & \hbox{\bf 1}
\\
$Q^3 A^{n - 1}$  & \hbox{\bf 1}
\\
$A \bar{U}^2$  & \asymm
\\
$\bar{U}^{N}$  & \hbox{\bf 1}
\\
$L^3$ &  \asymmthree
\end{tabular}
\eeq
The low energy (below $\Lambda_N$) degrees of freedom are
\beq\nonumber
\begin{tabular}{c|c|c}
 & $SU(3)$  & $SU(N)$  \\
\hline
$(Q\bar{U})$ & \fund & \fund
\\
$(S \bar{U})$ & {\bf 1} & \fund
\\
$(Q A^n)$ & \fund & \hbox{\bf 1}
\\
$(S A^n)$ & {\bf 1} & \hbox{\bf 1}
\\
$(Q^3 A^{n - 1})$ &  {\bf 1} & \hbox{\bf 1}
\\
$(S Q^2 A^{n - 1})$ & $\afund$ & \hbox{\bf 1}
\\
$(A \bar{U}^2)$ & \hbox{\bf 1} & \asymm
\\
$(\bar{U}^{N})$ & \hbox{\bf 1} & \hbox{\bf 1}
\\
L & $\afund$ & $\afund$
\end{tabular}
\eeq
The effective superpotential is given by the sum of the tree
superpotential and a dynamical superpotential.
\beq\bal
W_{\rm eff} &=
{{1}\over{\Lambda_N^{2N-1}}}
\biggl\{
\left[ (Q A^n) (Q \bar{U})^3 (S \bar{U})^2
+(S A^n) (S \bar{U})  (Q \bar{U})^2 \right] (A \bar{U}^2)^{n - 1} \\
&\qquad\qquad\quad
+ (Q^3 A^{n - 1}) (S \bar{U}) (A \bar{U}^2)^n
+ (S Q^2 A^{n - 1}) (Q \bar{U}) (A \bar{U}^2)^n
\\
&\qquad\qquad\quad
+ (\bar{U}^{N}) (S A^n) (Q^3 A^{n - 1})
+ (\bar{U}^{N}) (Q A^n) (S Q^2 A^{n - 1})\biggr\}
\\
&\qquad
+ \la L (Q\bar{U}).
\eal\eeq

Integrating out $L$ and $(Q \bar{U})$ leaves $SU(3)$ with one flavor
$(Q A^n)$ and $(S Q^2 A^{n-1})$ as well as some singlets with a
superpotential
\beq\bal
W_{\rm eff} &=
{{1}\over{\Lambda_N^{2N-1}}}
\biggl[
 (Q^3 A^{n - 1}) (S \bar{U}) (A \bar{U}^2)^n
+ (\bar{U}^{N}) (S A^n) (Q^3 A^{n - 1})
\\
&\qquad\qquad\quad
+ (\bar{U}^{N}) (Q A^n) (S Q^2 A^{n - 1})\biggr].
\eal\eeq
The last two terms in this superpotential are mass terms when the baryon
$(\bar{U}^N)$
has a VEV, so on this branch of the moduli space we find a dynamical
superpotential:
\beq
W_{\rm dyn} \sim (\bar{U}^N)^{1/3}
\eeq
Which forces the baryon to run away for any $N$.

Giving $N-m$ large VEVs to $\bar{U}$ and $S$ breaks $SU(N)$ down to
$SU(m)$, and the low energy theory is the original theory with $N$ replaced
by $m$, so the theory is still s-confining and the analysis goes through as
above.

For even $N$ ($N=2n$) and $k=2$ the low energy degrees of freedom are
\beq\nonumber
\begin{tabular}{c|c|cc}
 & $SU(2)$  & $SU(N)$ & $SU(2)$ \\
\hline
$(Q\bar{U})$ & \fund & \fund & {\bf 1}
\\
$(S \bar{U})$ & {\bf 1} & \fund &\fund
\\
$(A^n)$ & \hbox{\bf 1} & \hbox{\bf 1} & {\bf 1}
\\
$(Q^2 A^{n - 1})$ & {\bf 1}
& \hbox{\bf 1} & {\bf 1}
\\
$(S Q A^{n - 1})$ & \fund
& \hbox{\bf 1} & \fund
\\
$(S^2 A^{n - 1})$ & {\bf 1}
& \hbox{\bf 1} & {\bf 1}
\\
$(S^2 Q^2 A^{n - 2})$ & {\bf 1}
& \hbox{\bf 1} & {\bf 1}
\\
$(A \bar{U}^2)$ & \hbox{\bf 1} & \asymm & {\bf 1}
\\
$(\bar{U}^{N})$ & \hbox{\bf 1} & \hbox{\bf 1} & {\bf 1}
\\
$L$ & $\afund$ & $\afund$ & {\bf 1}
\end{tabular}
\eeq
The superpotential is
\beq\bal
W_{\rm eff} &\sim
{{1}\over{\Lambda_N^{2N-1}}}
\biggl\{
(A^n)(Q \bar{U})^2 (S \bar{U})^2 (A\bar{U}^2)^{n-2}
+(S^2 A^{n-1}) (Q \bar{U})^2 (A \bar{U}^2)^{n-1}
\\
&\qquad\qquad\quad
+ (Q^2 A^{n-1}) (S \bar{U})^2 (A \bar{U}^2)^{n-1}
+(S^2 Q^2 A^{n-2}) (A \bar{U}^2)^n\\
&\qquad\qquad\quad
+ (S Q A^{n-1}) (Q \bar{U}) (S \bar{U}) (A \bar{U}^2)^{n-1}
\\
&\qquad\qquad\quad
+ (\bar{U}^N) \left[ (A^n) (S^2 Q^2 A^{n-2}) + (Q^2 A^{n-1} )^2 \right.
\\
&\qquad\qquad\qquad\qquad\qquad
\left. + (SQ A^{n-1})  (SQ A^{n-1}) + (S^2 A^{n-1} )^2 \right]
\biggr\}
\\
&\qquad
+ \la L (Q\bar{U}).
\eal\eeq
Integrating out $L$ and $(Q \bar{U})$ leaves $SU(2)$ with one
flavor $(S Q A^{n - 1})$ and a superpotential
\beq\bal
W_{\rm eff} &\sim
{{1}\over{\Lambda_N^{2N-1}}}
\biggl\{
(Q^2 A^{n-1}) (S \bar{U})^2 (A \bar{U}^2)^{n-1} +
(S^2 Q^2 A^{n-2}) (A \bar{U}^2)^n
\\
&\qquad\qquad\quad
+ (\bar{U}^N) \left[ (A^n) (S^2 Q^2 A^{n-2}) + (Q^2 A^{n-1} )^2 \right.
\\
&\qquad\qquad\qquad\qquad\quad
\left. + (SQ A^{n-1} )  (SQ A^{n-1} ) + (S^2 A^{n-1} )^2 \right]
\biggr\}.
\eal\eeq
The last four terms are mass terms on the branch of moduli space
where $(\bar{U}^N)$ has a VEV, so gaugino condensation results
in the dynamical superpotential
\beq
W_{\rm dyn} \sim (\bar{U}^N)^{1/2} ~,
\eeq
which forces the baryon to run away for any $N$.

For even $N$ ($N=2n$) and $k=3$ the low energy degrees of freedom are
\beq\nonumber
\begin{tabular}{c|c|c}
 & $SU(3)$  & $SU(N)$  \\
\hline
$(Q\bar{U})$ & \fund & \fund
\\
$(S \bar{U})$ & {\bf 1} & \fund
\\
$(A^n)$ & \hbox{\bf 1} & \hbox{\bf 1}
\\
$(Q^2 A^{n - 1})$ & $\afund$
& \hbox{\bf 1}
\\
$(S Q A^{n - 1})$ & \fund
& \hbox{\bf 1}
\\
$(S Q^3 A^{n - 2})$ & {\bf 1}
& \hbox{\bf 1}
\\
$(A \bar{U}^2)$ & \hbox{\bf 1} & \asymm
\\
$(\bar{U}^{N})$ & \hbox{\bf 1} & \hbox{\bf 1}
\\
$L$ & $\afund$ & $\afund$
\end{tabular}
\eeq
The superpotential is
\beq\bal
W_{\rm eff} &=
{{1}\over{\Lambda_N^{2N-1}}}
\biggl\{
(A^n)(Q \bar{U})^3 (S \bar{U}) (A\bar{U}^2)^{n-2}
+ (Q^2 A^{n-1}) (S \bar{U})^2 (A \bar{U}^2)^{n-1}
\\
&\qquad\qquad\quad
+ (S Q A^{n-1}) (Q \bar{U}) (S \bar{U}) (A \bar{U}^2)^{n-1}
+(S Q^3 A^{n-2}) (A \bar{U}^2)^n
\\
&\qquad\qquad\quad
+ (\bar{U}^N) \left[ (A^n) (S Q^3 A^{n-2}) + (Q^2 A^{n-1} )^2 \right.
\\
&\qquad\qquad\qquad\qquad\quad
\left. + (SQ A^{n-1} )  (SQ A^{n-1} ) \right]
\biggr\}
\\
&\quad
+ \la L (Q\bar{U}).
\eal\eeq
Integrating out $L$ and $(Q \bar{U})$ leaves $SU(3)$ with one
flavor $(Q^2 A^{n - 1})$, $(S Q A^{n - 1})$ some singlets and a superpotential
\beq\bal
W_{\rm eff} &=
{{1}\over{\Lambda_N^{2N-1}}}
\biggl\{
 (Q^2 A^{n-1}) (S \bar{U})^2 (A \bar{U}^2)^{n-1}
+(S Q^3 A^{n-2}) (A \bar{U}^2)^n
\\
&\qquad\qquad\quad
 + (\bar{U}^N) \left[ (A^n) (S Q^3 A^{n-2}) + (Q^2 A^{n-1} )^2 \right.
\\
&\qquad\qquad\qquad\qquad\quad
\left. + (SQ A^{n-1} )  (SQ A^{n-1} ) \right]
  \biggr\}.
\eal\eeq
The last three terms are mass terms on the branch of moduli space
where $(\bar{U}^N)$ has a VEV, so gaugino condensation results
in the dynamical superpotential
\beq
W_{\rm dyn} \sim (\bar{U}^N)^{1/3} ~,
\eeq
which forces the baryon to run away for any $N$.

For $F=5$ and $N$ sufficiently large, the $SU(N)$ theory has an infrared
fixed point with a dual gauge group $SU(2) \times SU(2)$ \cite{fiveeasy}.
However for even $N$ it is known that the baryon
operator maps to a product of fields with scaling dimension less than 3,
so for $k \ge 3$ the baryon should again be forced to have a non-zero
SUSY breaking VEV.  For larger values of $F$ we expect that the $SU(N)$
dynamics have an infrared fixed point up to the point where asymptotic freedom
is lost ($F=2N+3$); unfortunately, little or nothing is know about the dual
descriptions
of these theories.


\end{document}